\documentstyle[epsfig]{mn}

\def\lesssim{\mathrel{\hbox{\rlap{\hbox{\lower4pt\hbox{$\sim$}}}\hbox{$<$}}}}
\def\gtrsim{\mathrel{\hbox{\rlap{\hbox{\lower4pt\hbox{$\sim$}}}\hbox{$>$}}}}
\title{\bf The Effect of Formation Redshifts on the Cluster
Mass--Temperature Relation}
\author[B. F. Mathiesen]{Benjamin F. Mathiesen \\ Department
of Physics, Stanford University, Stanford, CA 94305-4060 USA}

\begin{document}

\maketitle

\begin{abstract}
I employ an ensemble of hydrodynamical simulations and the
{\small XSPEC MEKAL} emission model to reproduce observable
spectral and flux-weighted temperatures for 24 clusters. 
Each cluster is imaged at 16 points in its history, which allows the
investigation of evolutionary effects on the mass-temperature relation.
In the zero redshift scaling relations, I find no evidence for
a relationship between cluster temperature and formation epoch
for those clusters which acquired 75\% of their final mass since
a redshift of 0.6. This result holds for both observable
and intrinsic intracluster medium temperatures, and implies that
halo formation epochs are not an important variable in
analysis of observable cluster temperature functions.
\end{abstract}

\begin{keywords}
galaxies: clusters: general --
large-scale structure of Universe
\end{keywords}

\section{Introduction}

Galaxy clusters are the youngest and largest organized structures
known to exist, and their ensemble properties can shed light
on many cosmological problems.
They arise from significantly overdense regions on cosmological
scales, which are exponentially rare events in a Gaussian
initial perturbation spectrum. The shape and normalisation of the
cluster mass function are therefore extremely sensitive to the
statistical properties of the primordial density field.
The evolution of cluster number densities is also tightly coupled
to the cosmic expansion rate, and can precisely constrain
$\Omega_m$. Much effort has been expended towards a measurement
of the cluster mass function and its evolution (see Henry
2000 for a summary of recent work in this area), but much remains
to be understood about the clusters themselves before we can be
confident of such results.


The properties of the local mass function can be constrained by using
either standard Press-Schecter (1974) (also Bond et al. 1991,
Lacey \& Cole 1993) theory, a more sophisticated analytical
model of the cosmic mass function (e.g. Sheth, Mo, \& Tormen 1999),
or fitting formula based on large-scale structure simulations (Jenkins et
al. 2000). These methods predict the number density of dark matter
haloes as a function of mass and redshift. Some relationship between
model variables (i.e., the total mass within some density threshold)
and a more easily observed quantity (e.g. X--ray temperature,
X--ray luminosity, or weak lensing mass) must therefore be assumed to
match cosmological predictions to the results of cluster surveys.
Intracluster medium (ICM) X--ray temperatures show particular promise
in this regard, since they demonstrate a tight ($\lesssim 20\%$
scatter) correlation with cluster mass 
components in both simulations (Mathiesen \& Evrard 2001,
total mass) and observations (Mohr, Mathiesen, \&
Evrard 1999, ICM mass), and the cluster X--ray temperature
function (XTF) currently
provides the tightest constraints on the cluster mass function.
It therefore seems likely that our best structure formation measurements
of $\Omega_m$ will rely on medium-redshift measurements of the XTF,
and it is essential to make sure that our interpretation of cluster
temperature functions is correct.

A modified form of the mass function allowing for differences between
a cluster's formation redshift and observed redshift was first proposed
by Kitayama \& Suto (1996), and has begun to be commonly implemented in
deriving constraints on the power spectrum normalization $\sigma_8$
at low redshfits (Kitayama \& Suto 1997, Kay \& Bower 1999,
Viana \& Liddle 1999). This extension to the theory produces
little change in the shape of the predicted mass function, but can
have an appreciable
effect on the predicted temperature function if one assumes that clusters
scatter around a mass-temperature scaling relation appropriate to their
formation redshift. This implies that
clusters have objectively identifiable formation events,
during which the majority of a cluster's mass coalesces for the
first time and the ICM virialises at a temperature
appropriate to the halo mass and formation epoch. These assumptions
are reasonable in the framework of standard Press-Schechter theory, but
perhaps pay too little heed to the more complicated, hydrodynamical
evolution of the ICM. 

The goal of this paper is to investigate the evolution of intrinsic
and observable cluster temperatures using an ensemble of 24
hydrodynamic cluster simulations, and test the appropriateness of
this extension to the interpretation of cluster temperature functions.
In the following section I describe the ensemble of simulated clusters
and our model for the ICM X--ray emission. In section 3, I look for
evidence of temperature evolution in our ensemble. Section 4 summarizes
the results of this letter. The Hubble constant is parameterized as
$H_0 = 100h$ km s$^{-1}$ Mpc$^{-1}$.

\section{Simulated and Observable Temperatures}

We use an ensemble of 24 hydrodynamical cluster simulations, divided
evenly between two reasonable cold dark matter (CDM) cosmological
models.  These models are OCDM ($\Omega_0 = 0.3$, $\sigma_8 =
1.0$, $h = 0.8$, $\Gamma = 0.24$) and $\Lambda$CDM ($\Omega_0 =
0.3$, $\lambda_0 = 0.7$, $\sigma_8 = 1.0$, $h = 0.8$, $\Gamma =
0.24$).  Here $\sigma_8$ is the linearly evolved, present day power
spectrum  normalization on $8h^{-1}$ Mpc scales. The initial conditions
are Gaussian random fields consistent with a CDM transfer function with
the specified $\Gamma$  (e.g. Bond \& Efstathiou 1984). The baryon
density is set in each case to $20\%$ of the total mass density
($\Omega_b = 0.2\Omega_0$).  The simulation scheme is P3MSPH: the
first stage is a P$^3$M (dark matter only) simulation to find cluster
formation sites in a large volume, and the second stage resimulates the
formation of individual clusters with higher resolution. Smoothed
particle hydrodynamics (SPH) is included in the individual cluster
simulations to resolve the ICM structure in detail.  The baryonic
component is modeled with $32^3$ particles, providing a typical mass
resolution of 0.01\% within the virial radius. The resulting
sample covers a little more than a decade in mass, ranging from
about $10^{14}$ to $3 \times 10^{15} M_{\odot}$. These simulations were
first presented a paper describing the cluster size-temperature relation
(Mohr \& Evrard 1997).

The simulations model the dynamical and thermodynamical effects of
gravitation, shock heating and adiabatic work on the ICM.  Several
potentially important pieces of physics are neglected.  Radiative
cooling is perhaps the most significant; our clusters cannot produce
cooling flows in their cores. Cooling flows have the potential to
greatly influence ICM luminosity and temperature measurements,
but the energy lost due to radiation is small compared to that released
in the process of gravitational collapse. We therefore expect that
the results of these simulations are comparable to observational
results which attempt to empirically account for the presence of cooling flows,
either through excision or explicit modeling of excess core emission.
Other neglected processes include galaxy feedback (Metzler \&
Evrard 1994) which can produce abundance
gradients and shallower gas profiles; preheating of the ICM
(Cavaliere, Menci, \& Tozzi 1998; Lloyd-Davies, Ponman, \& Cannon 2000),
which can raise the ICM entropy and limit the density of baryonic cores;
and electron temperature lag, which slightly cools X--ray
spectra in rich clusters (Chi\`{e}ze, Alimi \& Teyssier 1998;
Takizawa 1999) relative to the ion temperature. 
Further discussion of these issues in terms of
their relevance to cluster simulations can be found in Mathiesen
\& Evrard (2001, hereafter ME01) and Bialek, Evrard, \& Mohr (2001).

In ME01 we introduced an ensemble of spectrally and spatially resolved
X--ray surface brightness images derived from these simulations. We
used the {\small MEKAL} (Mewe, Lemen, \& van den Oord 1986) program
from the {\small XSPEC} utility, since this is the emission model most
commonly used to interpret observed ICM spectra. Each SPH
particle was assigned a 0.3 solar metallicity spectrum scaled
to its density and thermodynamic temperature and binned in 50 eV
intervals over the [0.1,20] keV bandpass. The clusters were
then ``observed'' by collecting photons in a circular window
centered on the minimum potential of the cluster, producing a combined
spectrum which incorporates emission from gas with a broad
distribution of phases.  We produced
{\em Chandra}-like combined spectra and spectral images by adopting
150 eV bins, a 0.5-9.5 keV bandpass, and finally convolving the
photon distribution with {\em Chandra's} effective area function and
a moderate ($3.4 \times 10^{20}$ cm$^{-2}$) absorbing column density.  
The physical scale of the observation windows used in this paper
varies from cluster to cluster, but usually corresponds to a fixed
mean interior overdensity of 500 times the critical density appropriate
to the redshift and cosmological model. This radius is labelled $r_{500}$,
and is a fixed fraction of the virial radius in any cosmology. The
radius $r_{200}$ is also used in this work, but corresponds less
closely to observable regions of the ICM.

The resulting spectra are surprisingly similar in character to isothermal
spectra with a temperature typically 10-20\% lower than the mass-weighted
mean thermodynamic ICM temperature. A semantic separation of the
two concepts appears to be necessary, so hereafter we refer to the
{\em Chandra}-like temperature just described as a {\em spectral}
temperature, $T_s$.  The mass-weighted mean thermodynamic temperature
will be referred to as the {\em virial} temperature $T_v$, because this
measure is found to precisely follow the virial relation
$M_{\rm tot} \propto T^{1.5}$ in these and other simulations.
$T_v$ is identical to the mass-weighted temperature $T_m$ defined in ME01.
When spectral fitting is limited to the 2.0-9.5 keV band
(similar to most published temperature determinations) the deviation
$\delta T_s$ between spectral and virial temperatures follows the relation
\begin{equation} 
\delta T_s \equiv \frac{T_v-T_s}{T_s}=
(0.19\pm 0.06)\log_{10}T_s [{\rm keV}]-(0.02\pm 0.04).
\label{dtsvst}
\end{equation}
This deviation steepens the observable mass-temperature relation
to $M_{\rm tot} \propto T_s^{1.62}$,
and implies that rich clusters are more massive than
their spectral temperatures would lead us to believe. The
difference between spectral and virial temperatures arises partly
from the presence of recent minor mergers in the gas column,
and partly from global ICM temperature profiles.  Both
effects contribute cooler gas to the observation window
and an overabundance of soft, line-emission photons to the
spectrum. I refer the reader to ME01 for further details on
this work.

The temperature bias described in Equation~\ref{dtsvst} has a
potentially important effect on any interpretation of the 
cluster temperature function and its evolution. As has already
been stated, these simulations do not include radiative cooling
and are therefore directly comparable only to X--ray data which
has accounted for the presence of cooling flows through excision
or explicit modeling of the excess emission. Since it has not been
possible until very recently to measure the spatial extent of
cooling flows in high-redshift clusters, most studies of cluster
evolution have evaluated the temperature and luminosity functions
at low redshifts without attempting such corrections. In such works,
the additional scatter introduced into cluster scaling relations by cooling
flows has been accepted as a source of uncertainty in the cosmological
constraints.

With the advent of {\em Chandra} and {\em XMM}, however, we can
do better.  Markevitch (1998) has shown that by excising the
central $50 h^{-1}$ Mpc of nearby clusters and including a 
cooling flow component in the core spectrum, the scatter in
the luminosity-temperature relation can be reduced by a factor
of two.  This allows for a more robust calculation of the maximum
observable volume for each cluster as a function of temperature,
as well as providing a cleaner estimate of the ICM temperature
in cooling flow clusters. The new telescopes should allow similar
corrections to be made in high-redshift clusters.  In order to achieve
precise cosmological constraints from a measurement of evolution in the
XTF, it is therefore desirable to model such cooling flow-corrected
temperatures.

Temperatures derived from high-quality {\em ASCA}
data (Markevitch et al. 1998) are commonly used in constructing
the local XTF or measuring the slope of cluster scaling relations.
These are flux-weighted mean spectral temperatures, which are biased
towards regions containing dense and/or cool gas.
This measure was found by Markevitch to produce a significant shift
in cooling flow cluster temperatures relative to an isothermal fit
to the combined spectrum. Other cluster temperatures,
on the other hand, were unchanged.

Such temperatures are difficult to reproduce precisely in the simulations,
mainly because the spatial extent and subdivision of {\em ASCA} images
is different for each cluster in the sample (Markevitch et al. 1998) The
large scale of {\em ASCA} spectral regions implies that the
temperatures in each
pixel are similar in character to the spectral temperatures described
earlier, but with more weight given to the luminous core than the
outer regions.  I simulate these temperatures by dividing our
observation windows into nine sectors with a morphology typical for
{\em ASCA} clusters: a core region with radius $r_{500}/4$;
an inner annulus surrounding this region with outer radius
$5r_{500}/8$, and an outer annulus surrounding this region
with outer radius $r_{500}$. The two annuli are each divided
into 4 quadrants. The average spectral temperature $T_s$ over all
nine sectors, weighted by total energy flux, will be referred to as
a {\em flux-weighted} temperature $T_f$ for the remainder
of this paper.  This definition of $T_f$ should not be confused
with the emission-weighted temperature $T_e$ described in ME01,
which is the density-weighted average thermodynamic temperature
calculated over a spherical volume. $T_f$ is, however, closely
comparable to the emission-weighted temperatures described in
the Markevitch et al. (1998) analysis.

The deviation $\delta T_f$ has a slope similar to that
of $\delta T_s$ (Equation \ref{dtsvst}), but a different normalisation:
\begin{equation} 
\delta T_f \equiv \frac{T_v-T_f}{T_f}=
(0.22\pm 0.05)\log_{10}T_f [{\rm keV}]-(0.11\pm 0.03).
\label{dtfvst}
\end{equation}
$T_f$ is generally higher than $T_s$ because of the extra weight given
to the core, but the scale-dependence of its deviation is similar.
The mean value of $\delta T_f$ ranges from about -10\% for poor clusters
to +10\% for rich clusters with virial temperatures on the order
of 10 keV.  Calculating the flux-weighted mean temperature
for a coarse grid does not free temperature measurements from a
spectral bias due to the presence of multiple phases in a gas column,
but it provides a more accurate estimate of the virial temperature for
rich clusters.

\begin{figure}
\epsfxsize 3.3in \epsfysize 3.3in \epsfbox{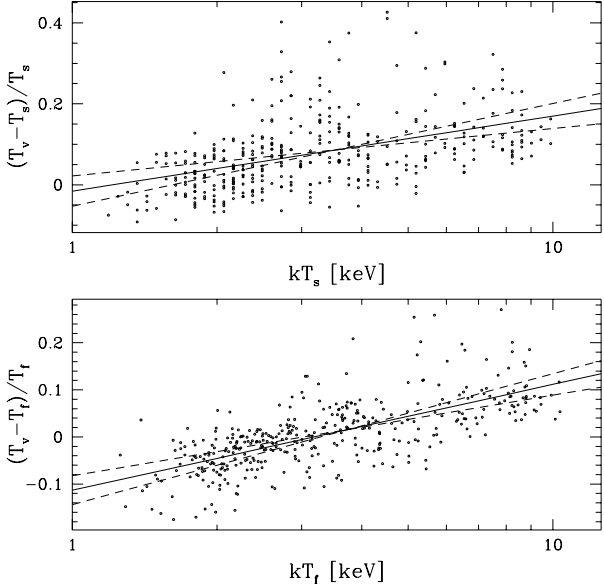}
\caption{The upper panel displays the deviation $\delta T_s$
between virial temperature and spectral temperature as a
function of the spectral temperature.  384 points are displayed,
corresponding to 16 evolutionary epochs of 24 independent
cluster simulations.  The best-fitting correlation between
$\delta T_s$ and $T_s$ (Equation 1) is independent of redshift.
The lower panel displays the corresponding relationship
between $\delta T_f$ and $T_f$, as reported in Equation 2. The
dashed lines in each panel display one-sigma variations on 
the best-fitting correlations.}
\label{dtfig}
\end{figure}

The deviations between virial temperatures and observable spectral
temperatures described in Equations \ref{dtsvst} and \ref{dtfvst}
do not vary with cosmological model or evolutionary epoch, nor do
they vary significantly when the radius of the observed region
is increased to enclose an mean overdensity of 200 times the
critical density. $\delta T_s$ and $\delta T_f$ are plotted agains
$T_s$ and $T_f$ in Figure \ref{dtfig}. These are robust measures of
an observational bias arising from realistic density and
temperature distributions in the ICM.  SPH simulations have
been shown to accurately reproduce
the large-scale morphology of real clusters (Mohr et al. 1995),
and a similar analysis of this particular ensemble
displayed an even closer structural correspondence (Mohr \& Evrard,
private communication). N-body simulations of dark matter evolution have
likewise been shown to produce merger histories which are in good
agreement with Press-Schechter theory (Lacey \& Cole 1994), so
it is likely that these variations are similar in magnitude to those
in real clusters.  Analysis of Eulerian simulations reveals a similar
level of clumping in the ICM (Bryan \& Norman, private communication).

\section{Modeling the Mass Function}

When converting a model mass function
$n(M,z)$ to an observable XTF $n(T,z)$, it is generally assumed that
the clusters formed at their observed redshift. Kitayama \& Suto
(1996) were the first relax this assumption in modeling the XTF,
applying the techniques of Lacey \& Cole (1993) to calculate
realistic distributions of formation epochs for the clusters at a
given redshift. They reasoned that clusters which formed
significantly earlier than their observation epoch would
be hotter than clusters of a similar mass which collapsed more
recently. Cosmological scaling of the background mean density and
temperature predicts a normalisation evolution in the mass-temperature
relation proportional to $h(z)/h_0 = \sqrt{\Omega_{\rm m}(1+z)^3 +
\Omega_\Lambda}$ for a flat universe (e.g. Bryan \& Norman 1998),
and the temperature corrections for such clusters can therefore be
rather large. Other groups have recently begun to take up this
standard in modeling the temperature function and its evolution
(Viana \& Liddle 1999, Kay \& Bower 1999), but this extension to
the model is not yet justified by observations.

Making this correction implies an additional assumption, however:
that the X--ray luminous regions of clusters are approximately
relaxed.  Merger events have the potential to significantly alter
a cluster's temperature when they occur, and they need not be very
large to do so (Cavaliere et al. 1999, ME01).  For
a cluster which formed at high redshift to maintain a temperature
appropriate for that epoch, it should have already accumulated most
of its observed mass.  The rate of mergers observed in simulations 
makes this scenario seem unlikely, although Kitayama and Suto cite
results from Eulerian simulations (Bryan \& Norman 1998) indicating
that a cluster's temperature doesn't change much after its formation.
We note that the temperature which they refer to is a simulation's 
luminosity-weighted temperature (similar to the $T_e$ used in ME01),
which is very similar to the core
temperature of the gas. Observational measures of the temperature
such as $T_s$ and $T_f$ are more heavily influenced by cool gas in
the outer regions of the cluster, and more susceptible to
minor merger events.

\begin{figure}
\epsfxsize 3.3in \epsfysize 3.3in \epsfbox{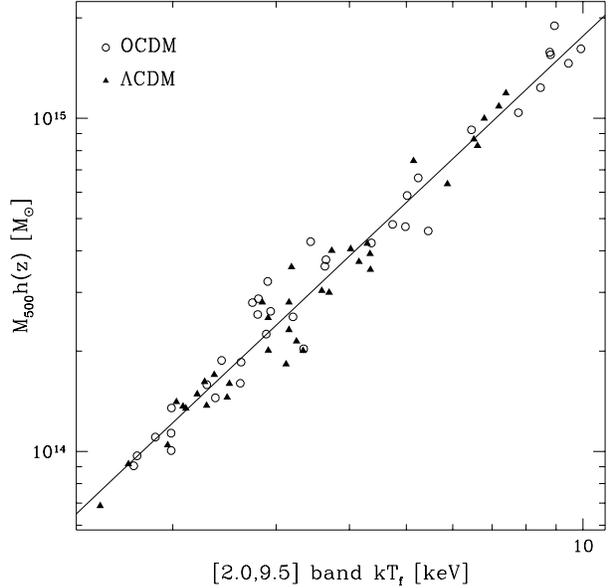}
\caption{The relationship between flux-weighted temperature $T_f$ and
total mass measured within $r_{500}$. Clusters at redshifts 0.0, 0.5
and 1.0 are included in the plot, with masses scaled by the cosmological
evolution factor h(z). The best-fitting relation is drawn, and its
parameters are reported in Equation \ref{tfbestfit}.}
\label{fluxmasstemp}
\end{figure}

These simulations can be used to test the sensitivity of observable
and virial temperatures to ongoing minor mergers. The mass of our
clusters as a function of flux-weighted temperature is plotted in
Figure \ref{fluxmasstemp}, and presents a tight correlation with only
18\% scatter:
\begin{equation}
\log_{10}(M_{\rm tot}h(z)) = (1.66\pm 0.04)\log_{10}T_f + (13.59\pm 0.02).
\label{tfbestfit}
\end{equation}
This plot combines cluster outputs at redshifts $z=1.0$, 0.5, and 0.
The small degree of scatter in this plot strongly implies that there is no
significant contamination of the temperature ensemble by clusters which
virialized early; when scaled to a similar epoch the three relations
$M(T_f;z)$ are identical. We can, however, probe this issue more deeply. 

\begin{figure}
\epsfxsize 3.3in \epsfysize 3.3in \epsfbox{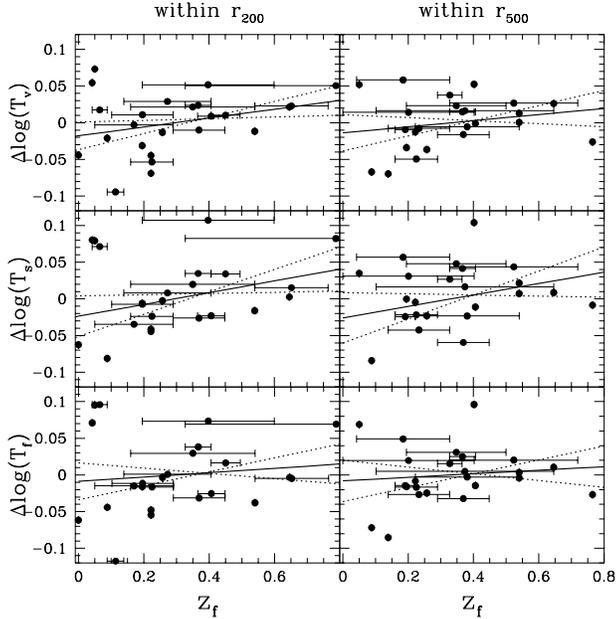}
\caption{The best-fitting mass-temperature relationship $\bar{T}(M)$
is calculated for the 24 clusters in this ensemble at a redshift of
zero, and the deviation of individual clusters from this relation
$\Delta\log T \equiv log(T/\bar{T})$ is plotted against their
formation redshift $z_f$. Three definitions of the temperature
(virial, spectral, and flux-weighted) and two observation windows
are examined. If the temperature of a cluster were strongly
influenced by its formation epoch, we would expect to see a
correlation between $\Delta\log T$ and $z_f$. The evidence for
correlation in this data set is marginal at best, and negligible
for the flux-weighted temperatures which are most similar
to {\em ASCA} measurements. Although
the simulated clusters display the expected cosmological evolution in
the normalization of their mass-temperature relation, the scaling
relation at zero redshift does not appear to be contaminated
by clusters which virialized at earlier epochs.}
\label{tvszform}
\end{figure}

We define a cluster's formation epoch $z_f$ as the redshift at which it has
acquired $75\% \pm 7.5\%$ of its final mass, following the convention
used by Viana \& Liddle (1999) in their recent paper constraining
cosmological parameters. If the extension of Kitayama \& Suto is relevant
to our interpretation of the local XTF, then we should see a correlation
between cluster temperature and $z_f$: clusters with high formation
redshifts should, on the average, have temperatures higher than the mean
mass-temperature relation.  Figure \ref{tvszform} plots this difference
for virial, spectral, and flux-weighted temperatures within observation
windows of radius $r_{500}$ and $r_{200}$.  Error bars
along the $z_f$ axis are given to objects which passed through the mass
threshold with a significant component of continuous accretion, so that
the cluster had between 67.5\% and 82.5\% of its final mass in
more than one output frame.  The uncertainties implied by these
error bars are not used in the statistical analysis. All the
best-fitting lines for the six data sets are consistent with zero slope,
although it is
fair to say that there is evidence for a slight positive correlation. 
On the other hand, this correlation is largely driven by the rare
clusters which formed at very high redshifts; most of these
objects formed at $z_f < 0.6$ and are evenly distributed about the mean
mass-temperature relationship.

Correlation
coefficients and best-fit line parameters for these data
are summarized in Table \ref{tzcorrtable}.  The correlations within
$r_{200}$ are significantly influenced by a cluster with a formation
redshift of 0.8 and very large error bars on that value. This cluster
acquired most of its mass very early in the simulation and grew through
gradual accretion thereafter, so the reshift range during which it had
a mass of $75\% \pm 7.5\%$ its final mass is very long.  It's temperature,
while higher than the ensemble average, is well within the variations
seen for more recently formed clusters.  If this point is left out
the analysis, then all the slopes become consistent with
zero in their one-sigma uncertainty range, and the correlation
coefficients drop to 0.24 ($T_v$), 0.19 ($T_s$),
and 0.005 ($T_f$). These coefficients correspond respectively 
to 32\%, 39\%, and 100\% probabilities of uncorrelated data.

\begin{table}
\begin{center}
\begin{tabular}{llcccc}
\multicolumn{2}{l}{Temperature} & $R$ & slope & intercept \\ \hline
$r_{200}$ & $T$   & 0.31(15\%) & $0.060\pm 0.044$ & $-0.018\pm 0.017$ \\
          & $T_s$ & 0.28(20\%) & $0.080\pm 0.066$ & $-0.024\pm 0.025$ \\
          & $T_f$ & 0.12(58\%) & $0.030\pm 0.060$ & $-0.009\pm 0.022$ \\
$r_{500}$ & $T$   & 0.21(33\%) & $0.041\pm 0.051$ & $-0.014\pm 0.020$ \\
          & $T_s$ & 0.28(20\%) & $0.078\pm 0.086$ & $-0.026\pm 0.034$ \\
          & $T_f$ & 0.11(68\%) & $0.024\pm 0.057$ & $-0.008\pm 0.022$ \\ \hline
\end{tabular}
\end{center}
\caption{A correlation analysis of the data presented in Figure
\ref{tvszform}. $R$ is the correlation coefficient of the data,
and is also translated into the probability that an uncorrelated
set of 24 random points would produce a correlation coefficient
at least that large.  The slope and intercept of the best-fitting lines
plotting in Figure \ref{tvszform} are also given.}
\label{tzcorrtable}
\end{table}

The discrepancy between our intuition (clusters which first virialized
at an early epoch should be hotter) and these simulations can be resolved
by acknowledging the essentially dynamic nature of clusters in a
low-density universe.  Multiple lines of observational
evidence point to an $\Omega_m \sim 0.3$ cosmology in which
clusters are still forming at the present day, and the theoretical
construct of a relaxed, virialized cluster seems to have few counterparts
in the observable population. Rather than treating clusters as static
fossils of the primordial density field, we should attempt to model them
explicitly as evolving entities.  One example of such a
model has been presented by Cavaliere et al. (1999), who
analyze ICM structure in terms of ``punctuated equilibrium'', a sequence of
merger shocks followed by partial relaxation of the ICM to the shock
boundary conditions.  Their model agrees with these simulations
in showing that minor merger events should have a more important
influence on the evolution of ICM temperatures than major mergers,
and makes some important predictions about the behavior
of ICM scaling relations for rich groups and poor clusters.

\section{Conclusions}

An analysis of the evolutionary history of simulated clusters
shows only a slight dependence of cluster temperature on formation
epoch out to $z_{\rm f} \sim 0.6$, although the small number of
clusters in this ensemble leaves open the possibility of evolution
in the temperatures of clusters which
formed at high redshifts. It is also worth reiterating that when
using a flux-weighted temperature $T_f$, which is the definition most
appropriate to analysis of the cluster temperature function, this
correlation essentially vanishes. This lack of dependence on formation
epoch can be traced to a high frequency of smaller (mass ratio 
$\lesssim 25\%$) merger events, which introduce cool gas into
the ICM and allow it to approach an equilibrium appropriate to
the merger epoch. This work
implies that extending analysis of the temperature function
to account for the difference between a cluster's formation
and observation redshifts is not necessary in a dynamically young
halo population.

\vspace{-0.15in}
\section*{Acknowledgments}
I would like to thank Joseph J. Mohr, Keith L. Thompson, and
Jeffrey A. Willick for valuable discussions which greatly aided
this work. This research was supported by a Terman Fellowship
and the Research Corporation.


\vspace{-0.15in}
\begin{thebibliography}{99}
\bibitem{bialek} Bialek J. J., Evrard A. E., \& Mohr J. J., 2001, preprint
(astro-ph/0010584)
\bibitem{bond84} Bond J. R. \& Efstathiou G., 1984, ApJL, 285, L45
\bibitem{bond91} Bond J. R., Cole S., Efstathiou G., \& Kaiser N., 1991,
ApJ, 379, 440
\bibitem{bryan} Bryan G. L. \& Norman M. L., 1998, ApJ, 495, 80
\bibitem{cav98} Cavaliere A., Menci N., \& Tozzi P., 1998, ApJ, 501, 493
\bibitem{cav99} Cavaliere A., Menci N., \& Tozzi P., 1999, MNRAS, 308, 599
\bibitem{chieze} Chi\`{e}ze J.-P., Alimi J.-M. \& Teyssier R., 1998, ApJ,
495, 630
\bibitem{henry} Henry, J. P., 2000, ApJ, 534, 565
\bibitem{jenkins} Jenkins A., Frenk C. S., White S. D. M., Colberg J. M.,
Cole S., Evrard A. E., \& Yoshida N., 2000, preprint (astro-ph/0005260)
\bibitem{kay} Kay S. T. \& Bower R. G., 1999, MNRAS, 308, 664
\bibitem{kita96} Kitayama T. \& Suto Y., 1996, ApJ, 469, 480
\bibitem{kita97} Kitayama T. \& Suto Y., 1997, ApJ, 490, 557
\bibitem{lacey93} Lacey C. \& Cole S., 1993, MNRAS, 262, 627
\bibitem{lacey94} Lacey C. \& Cole S., 1994, MNRAS, 271, 676
\bibitem{lloyd} Lloyd-Davies E. J., Ponman T. J. \& Cannon D. B., 2000,
MNRAS, 315, 689
\bibitem{m98b} Markevitch M., 1998, ApJ, 504, 27 
\bibitem{m98a} Markevtich M., Forman W. R., Sarazin C. L., \& Vikhlinin A.,
1998, ApJ, 503, 77
\bibitem{math01} Mathiesen B. F. \& Evrard A. E., 2001, ApJ, in press
(astro-ph/0004309)
\bibitem{metzler} Metzler C. A. \& Evrard A. E., 1994, ApJ, 437, 564
\bibitem{mewe} Mewe R., Lemen J. R., \& van den Oord G. H. J., 1986, A\&AS,
65, 511
\bibitem{mohr95} Mohr J. J., Evrard A. E., Fabricant D. G., \& Geller M. J.,
1995, ApJ, 447, 8
\bibitem{mohr97} Mohr J. J. \& Evrard A. E., 1997, ApJ, 491, 38
\bibitem{mohr99} Mohr J. J., Mathiesen B. F., \& Evrard A. E., 1999, ApJ,
517, 627
\bibitem{press} Press W. H. \& Schechter P., 1974, ApJ, 187, 425
\bibitem{sheth} Sheth R. K., Mo H. J., \& Tormen G., 1999, preprint
(astro-ph/9907024)
\bibitem{takizawa} Takizawa, M. 1999, ApJ, 520, 514
\bibitem{viana} Viana P. T. P. \& Liddle A. R., 1999, MNRAS, 303, 535

\end{thebibliography}
\end{document}